\documentclass{aa}
\usepackage{psfig}
\def\la{\;
\raise0.3ex\hbox{$<$\kern-0.75em\raise-1.1ex\hbox{$\sim$}}\; }
\def\ga{\;
\raise0.3ex\hbox{$>$\kern-0.75em\raise-1.1ex\hbox{$\sim$}}\; }
\begin{document}
\thesaurus{03(09.13.2; 02.12.3; 11.17.1; 11.17.4 I Zw 18)}
\title{
Argon and Silicon abundances in 
the damped Ly$\alpha$ system I~Zw~18
}
\author{Sergei A. Levshakov\inst{1}
\and Wilhelm H. Kegel\inst{2}
\and Irina I. Agafonova\inst{1}}
\offprints{S. A. Levshakov}
\mail{lev@astro.ioffe.rssi.ru}
\institute{
Department of Theoretical Astrophysics,
Ioffe Physico-Technical Institute, 
194021 St. Petersburg, Russia
\and
Insitut f\"ur Theoretische Physik der Universit\"at Frankfurt
am Main, 60054 Frankfurt/Main 11, Germany}
\date{Received 00 December 2000 / Accepted 00 December  2000}
\titlerunning{Argon and Silicon abundances in I~Zw~18}
\authorrunning{S.A. Levshakov et al.}
\maketitle
\begin{abstract}
We show that the difference between the Ar and Si relative abundance  
ratio derived from {\it FUSE} absorption spectra   
and from  the \ion{H}{ii} regions of I~Zw~18 is 
a consequence of the
microturbulent analysis applied to the absorption spectra.
{\it FUSE} observations were performed with a large
entrance aperture which fully covered the galaxy.
This means that the observed profiles are averaged over the full
body of I~Zw~18, 
implying that large-scale velocity fields influence the
absorption-line profiles.
Taking this into account, we show
that the absorption spectra are consistent with the same metal abundances
as those derived from the \ion{H}{ii} regions.
It follows that
no significant ionization correction as suggested by Izotov and
collaborators to describe metal contents in damped Ly$\alpha$
systems (DLA) 
is required to model abundances in 
the neutral gas of I~Zw~18 
(a local DLA system).
Using a mesoturbulent approach and 
applying the generalized radiative transfer
equation to the
\ion{Ar}{i}\,$\lambda1048$ and \ion{Si}{ii}\,$\lambda1020$ lines
observed by Vidal-Madjar et al., we found that 
the profiles may be reproduced with 
log (Ar/Si) $\simeq - 0.8$ and
$N(\ion{Si}{ii}) \simeq 4\times10^{15}$ cm$^{-2}$.

\end{abstract}

\section{Introduction}

The blue compact galaxy (BCG) I~Zw~18 (Mrk 116) 
has been an intensively studied
object for the last three decades since 
the first spectroscopic observations
by Zwicky (1966). 
I~Zw~18 shows an intense and recent burst of star formation which
makes this galaxy an attractive target for studies of 
star formation history. 
Recent observations revealed, for example, two 
stellar populations in I~Zw~18:
10--20 Myr red supergiants and 0.1--5 Gyr asymptotic
giant branch stars (\"Ostlin 2000).

Amongst BCGs, I~Zw~18 shows the lowest oxygen abundance~: the northwest (NW) and
the southeast (SE) bright \ion{H}{ii} regions yield log (O/H) $= -4.83 \pm 0.03$
and $-4.82 \pm 0.03$, respectively (Izotov et al. 1999). Relative to solar values,
one finds [O/H]\footnote{Using the customary
definition [X/H] = $\log [N({\rm X})/N({\rm H})] - 
\log [N({\rm X})/N({\rm H})]_\odot$, and solar abundances from Grevesse et al. (1996)
except for Ar for which the weighted average value $-5.48 \pm 0.04$ from
Sofia \& Jenkins (1998) is adopted.}
$ = -1.70 \pm 0.08$, i.e.
$Z/Z_\odot \simeq 1/50$. 
For argon abundances, Izotov et al. obtained log (Ar/H) $= -6.90 \pm 0.05$ (NW)
and a slightly lower ratio for the SE region, log (Ar/H) = $-7.23 \pm 0.05$.  
Silicon abundances measured by Izotov \& Thuan (1999) give  
log (Si/H) $= -6.29 \pm 0.22$ (NW) and log (Si/H) $= -6.30 \pm 0.22$ (SE).
Thus, these values imply for 
the NW emission patch
[Ar/H] = $-1.42 \pm 0.06$ and [Si/H] = $-1.85 \pm 0.22$.

\begin{figure*}
\vspace{-1.5cm}
\hspace{0.0cm}\psfig{figure=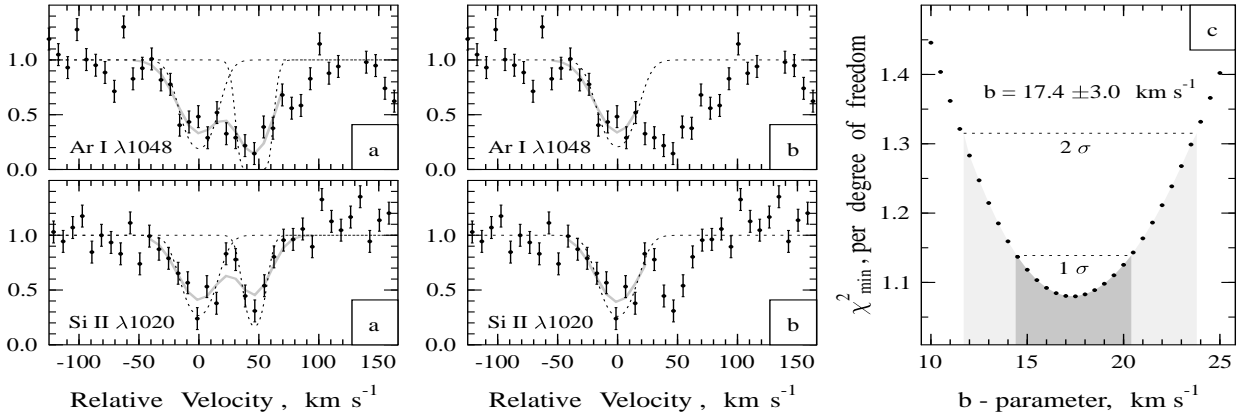,height=9.0cm,width=18.0cm}
\vspace{-2.0cm}
\caption[]{({\bf a}) -- Observed normalized intensities
(dots with error bars) vs radial velocities 
from {\it FUSE} spectra of \ion{Ar}{i} (top panel) and \ion{Si}{ii} (bottom panel)
obtained by Vidal-Madjar et al. (2000).
Smooth grey lines are the synthetic spectra (convolved with the spectrograph
function) of the two-component microturbulent model  
calculated from the $\chi^2$ minimization as described in the text.
Smooth dotted lines show the corresponding unconvolved spectra. 
Components at $\Delta v \simeq 49$ km~s$^{-1}$ are the Galactic H$_2$ lines
[\,L(4-0)P(1) in top panel and L(7-0)P(4) in bottom panel\,] from the high-velocity
cloud at $-160$ km~s$^{-1}$. 
({\bf b}) -- Same as {\bf a} but for a one-component microturbulent model.
({\bf c}) -- Confidence regions in the $\chi^2_{\rm min} - b$ plane
calculated from the simultaneous fit of the \ion{Ar}{i} and
\ion{Si}{ii} lines shown in panels {\bf b}. 
The parabola vertex corresponds to the best value of $b = 17.4$ km~s$^{-1}$
}
\end{figure*}

The neutral gas properties in I~Zw~18 
have been probed
in both the radio (see van Zee et al. 1998 and references cited therein)  
and the UV range 
(Kunth et al. 1994; Pettini \& Lipman 1995; Vidal-Madjar et al. 2000).
High velocity and high spatial resolution
radio observations have shown that the overall kinematics of the \ion{H}{i} gas
associated with I~Zw~18 is very complex and the neutral gas velocity dispersion 
$\sigma$ equals $13 - 14$ km~s$^{-1}$, or $b \equiv \sqrt{2}\sigma = 18 - 20$
km~s$^{-1}$  (van Zee et al. 1998).
The \ion{H}{i} column density in front of I~Zw~18 deduced from
the Ly$\alpha$ absorption profile by Kunth et al.,
$N(\ion{H}{i}) = (3.5 \pm 0.5)\times10^{21}$ cm$^{-2}$, is comparable to
the peak \ion{H}{i} surface density found by van Zee et al.,
$N(\ion{H}{i}) \simeq 3.0\times10^{21}$ cm$^{-2}$.
This means that the neutral gas in I~Zw~18 can be considered as a local damped
Ly$\alpha$ system (DLA) which is similar to high redshift DLAs observed in the
light of background quasars. 

First measurement of the O/H abundance in the \ion{H}{i} region have 
indicated a possible discrepancy between the metal content in the neutral
gas and in the \ion{H}{ii} regions (Kunth et al. 1994). This, however,  was not
confirmed in later studies by Pettini \& Lipman (1995)
and  by van Zee et al. (1998) who have shown that both the neutral and ionized  
gas in I~Zw~18 may have the same oxygen abundance.

New observations carried out with the {\it Far
Ultraviolet Spectroscopic Explorer} ({\it FUSE}) by Vidal-Madjar et al. (2000)
produce a similar puzzle~: the column
density ratio deduced from 
the \ion{Ar}{i}\,$\lambda1048$ and 
\ion{Si}{ii}\,$\lambda1020$ lines in the neutral gas, 
log~(Ar/Si) $= -1.32$, 
differs significantly from that observed in the \ion{H}{ii} regions,
log~(Ar/Si)$_{\rm NW} = -0.61 \pm 0.22$ and
log~(Ar/Si)$_{\rm SE} = -0.74 \pm 0.22$ 
(Izotov et al. 2000). Izotov et al. suggested that
ionization in DLAs may affect abundance ratios. 
To interpret the observations, 
they suggested a model consisting of two regions with {\it substantially different}
metal contents, i.e. implicitly recalling an idea of the existence of
two media in BCGs (a pristine gas, unprocessed
since the big bang and a gas polluted by nucleosynthetic products --
Kunth \& Sargent 1986).
In general, the discovery of such a primordial gas would be
of great importance for cosmology, as noted by Kunth et al. (1994).

In this Letter, we report on the study of the \ion{Ar}{i}\,$\lambda1048$ and 
\ion{Si}{ii}\,$\lambda1020$ profiles of I~Zw~18 
published by Vidal-Madjar et al. (2000).
We show that these lines can be modeled  
under the assumption that  
the metal content in the neutral gas is the same as
in the \ion{H}{ii} regions without refering to an
ionization correction.

\section{Data analysis and results}

Spectroscopic observations of I~Zw~18 with {\it FUSE} 
in the range $\sim 980 - 1187$ \AA\, are described in detail
by Vidal-Madjar et al. (2000).
The spectrum was obtained with a 
resolution of about $\lambda/\Delta\lambda \sim 10,000$
and a signal-to-noise
ratio of S/N $\sim 10$ per resolution element.
The large entrance aperture ($30'' \times 30''$)
fully covers the galactic surface ($\sim 10'' \times 4''$).
As noted by Vidal-Madjar et al., {\it FUSE} produces the
average absorption over the full body of the galaxy.
In this case the analysis of saturated absorption 
lines is not an easy and unambiguous task. 
The main difficulty
is connected with 
the line broadening by large-scale irregular (stochastic)
velocity fields. The influence of the finite correlation length
on the line profile depends on the details of the observation.
If one
considers the line formation process in the light of a point source,
then the observed spectrum reflects only one realization of the velocity
field and, hence, large deviations from the expectation value
of the intensity
$\langle I_\lambda \rangle$ may occur if the correlation length of the
velocity field $\ell$ is not very small compared to the size of the absorbing
region (Levshakov \& Kegel 1997).
If, however, the spectrograph aperture covers an essential part of the
galactic surface, then $\langle I_\lambda \rangle$ should reasonably
well correspond to the observations. 
But also in this case the standard Voigt-fitting analysis
(based on the assumption of microturbulence)
may
yield misleading results (Levshakov \& Kegel 1994).
Below we demonstrate this effect using the published {\it FUSE} data. 

\begin{figure}
\vspace{-1.4cm}
\hspace{0.0cm}\psfig{figure=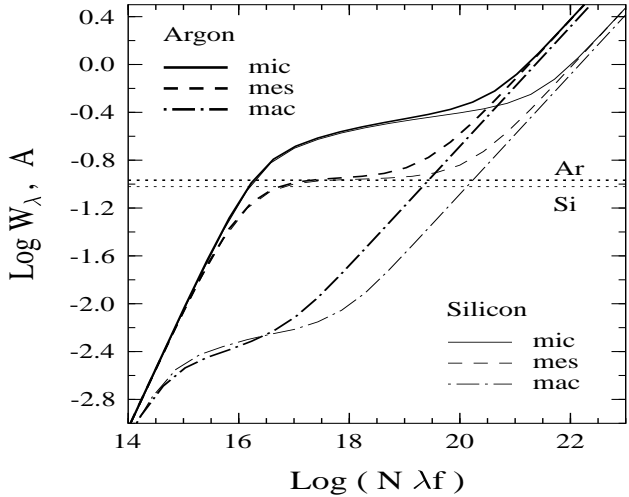,height=14.0cm,width=14.0cm}
\vspace{-6.0cm}
\caption[]{
Curves of growth for \ion{Ar}{i}\,$\lambda1048$ and \ion{Si}{ii}\,$\lambda1020$
calculated with the fixed values of the velocity dispersion 
$\sigma = 13$ km~s$^{-1}$ and the kinetic temperature $T_{\rm kin} = 200$~K.
Three types of lines correspond to micro- , meso- , and macroturbulence
($\ell = 0$, $L/\ell = 1.45$, $L/\ell = 0$, respectively).
Dotted horizontal lines mark the estimated values of the 
\ion{Ar}{i} and \ion{Si}{ii} equivalent widths 
}
\end{figure}

\subsection{A microturbulent approach}

We begin with the standard Voigt-fitting analysis which assumes a completely
uncorrelated velocity field, i.e. $\ell = 0$ ({\it microturbulence}).
In our example we use the published data from Fig.~1 of Vidal-Madjar et al.
to illustrate how 
the column density ratio deduced from two lines, \ion{Ar}{i} and \ion{Si}{ii},
is affected by the underlying assumptions.
From these data  
given in arbitrary units we have calculated normalized
intensities using the plotted synthetic profiles, 
and then added to the normalized data  $1\sigma$ error bars
corresponding to S/N = 10 (dots and error bars in Figs.~1{\bf a}, 1{\bf b}). 
Both profiles were
centered at $v = 0$ km~s$^{-1}$ according to the synthetic spectra of
Vidal-Madjar et al., and the internal uncertainty of our velocity scale 
calibration was estimated to be about $\pm 5$ km~s$^{-1}$.   
 
The oscillator strengths of the \ion{Ar}{i}\,$\lambda1048.2119$ \AA\, and
\ion{Si}{ii}\,$\lambda1020.6989$ \AA\, lines, $f_{\ion{Ar}{i}} = 0.257$
and $f_{\ion{Si}{ii}} = 0.01391$, were taken from Federman et al. (1992)
and from Charro \& Mart\'in (2000), respectively. 

The absorption lines of \ion{Ar}{i} and \ion{Si}{ii} are partly blended 
with the Galactic H$_2$ L(4-0)P(1) and L(7-0)P(4) lines, respectively.
To estimate 
the \ion{Ar}{i} and \ion{Si}{ii} column densities and the $b$-parameter,
we firstly 
fitted two component Voigt profiles to the observed intensities
via $\chi^2$ minimization. In this procedure the theoretical profiles
were convolved with the instrumental point-spread function which is supposed
to be a Gaussian with the width of 26 km~s$^{-1}$. The best fit with
$\chi^2_{\rm min} = 34.25$ ($M = 35$ data points, and $\nu = 28$ degrees of freedom) 
is shown in Fig.~1{\bf a} by grey solid curves which simultaneously mark 
the data points
involved in the optimization procedure.
Dotted curves in Fig.~1{\bf a} show the corresponding unconvolved profiles.
As seen from this figure,
the H$_2$ blends do not affect the cores of the 
\ion{Ar}{i} and \ion{Si}{ii} lines much. 
This is why in the further analysis we used
only the blue wings and the central parts of these lines. 

Our main results are shown in Figs.~1{\bf b} and 1{\bf c}. 
Fig.~1{\bf b} presents the best fit
with $\chi^2_{\rm min} = 18.36$ ($M = 19$, $\nu = 16$), 
$N_{\ion{Ar}{i}} = 6.83\times10^{13}$ cm$^{-2}$,
$N_{\ion{Si}{ii}} = 1.11\times10^{15}$ cm$^{-2}$, and 
$b = 17.4$ km~s$^{-1}$. Fig.~1{\bf c} shows the calculated $\chi^2_{\rm min}$
values as a function of $b$ with the $1\sigma$ and $2\sigma$ confidence
levels. Here, for each point we fixed the value of $b$
and minimize $\chi^2$ by varying all other parameters.

The estimated ratio
of log~(Ar/Si) $= -1.21$ 
is consistent within the uncertainty interval of\, 0.11 dex 
with the value  $-1.32$  cited by Izotov et al. (2000).
The most likely value for $b$, $17.4 \pm 3$ km~s$^{-1}$, is in an excellent
agreement with the radio observations of the \ion{H}{i} gas in I~Zw~18
carried out
by van Zee et al. (1998). The fact that $b_{\ion{H}{i}} \simeq b_{\ion{Ar}{i}}$
implies that the kinetic temperature of the neutral gas in I~Zw~18 is less
than 600~K and that the line broadening is caused mainly by turbulent
motions. 

To summarize this section we note that 
the microturbulent results lead to the puzzle mentioned above
which encouraged us to try
a more general model.

\begin{table}
\centering
\caption{Ar and Si abundances of neutral gas in I~Zw~18}
\label{tab1}
\begin{tabular}{lccccc}
\hline
\noalign{\smallskip}
$L/\ell$ & $N_{\ion{Si}{ii}}$, cm$^{-2}$ & Ar/Si$^\ddagger$ & 
$\frac{1}{\nu}\chi^2_{\rm min}$ & [Ar/H]$^\dagger$ & [Si/H]$^\dagger$  \\
\noalign{\smallskip}
\hline
\noalign{\smallskip}
 0   & 1.0(19) & $-1.93$ & 1.33 & $\;\;\;1.03$&$\;\;\;1.91$  \\
 0.50& 4.4(18) & $-1.82$ & 1.13 & $\;\;\;0.77$&$\;\;\;1.11$ \\
 0.75& 1.6(18) & $-1.57$ &1.08 & $\;\;\;0.58$&$\;\;\;1.11$ \\
 0.90& 4.1(17) & $-1.12$ &1.06 & $\;\;\;0.43$&$\;\;\;0.51$ \\
 1.00& 4.9(16) & $-0.32$ &1.06 & $\;\;\;0.31$&$-0.41$ \\
 1.10& 1.3(16) & $\;\;\;0.07$ & 1.07 & $\;\;\;0.14$ &$-0.98$ \\
 1.15& 9.5(15) & $\;\;\;0.10$ & 1.07 & $\;\;\;0.02$ &$-1.12$ \\
 1.20& 7.5(15) & $\;\;\;0.02$ & 1.07 & $-0.17$ &$-1.22$ \\
 1.25& 6.3(15) & $-0.05$& 1.08 & $-0.32$ &$-1.30$ \\
 1.30& 5.4(15) & $-0.34$& 1.08 & $-0.66$ &$-1.36$ \\
 1.35& 5.0(15) & $-0.71$& 1.08 & $-1.08$&$-1.40$ \\
 1.40& 4.4(15) & $-0.75$& 1.08 & $-1.17$&$-1.46$ \\
 1.45& 4.0(15) & $-0.84$& 1.08 & $-1.30$&$-1.50$ \\
 1.50& 3.7(15) & $-0.90$& 1.08 & $-1.39$&$-1.53$ \\
 2.00& 2.5(15) & $-1.09$ &     1.08 & $-1.76$&$-1.71$ \\
 5.00& 1.5(15) & $-1.19$ &     1.08 & $-2.09$&$-1.94$ \\
10.0 & 1.3(15) & $-1.20$ &     1.08 & $-2.17$&$-2.00$ \\
$\infty$ &1.1(15)&$-1.21$&     1.09 & $-2.23$&$-2.06$ \\
\noalign{\smallskip}
\hline
\noalign{\smallskip}
\multicolumn{6}{l}{$^\ddagger$\,Ar/Si $\equiv \log (N_{\ion{Ar}{i}}/N_{\ion{Si}{ii}})$ }\\
\multicolumn{6}{l}{$^\dagger$\, the internal accuracy of these values is $\simeq 0.1$ dex}
\end{tabular}
\end{table}

\subsection{A mesoturbulent approach}

21 cm observations with high spatial resolution show
that the line profile varies with position (van Zee et al. 1998).
This is a clear indication that large-scale motions determine the
line profiles and that the microturbulent approach is not well
founded. We therefore generalize our analysis to include
the effects of a finite correlation length,
i.e. $\ell \neq 0$ ({\it mesoturbulence}) and to study the limiting case 
$L/\ell \rightarrow 0$ ({\it macroturbulence}).

The simulation of the mesoturbulent 
\ion{Ar}{i} and \ion{Si}{ii} profiles has been carried out using the simplified 
model of a plane-parallel slab of geometrical size $L$ with homogeneous
turbulence and uniform kinetic temperature, $T_{\rm kin}$.
Spe\-cifying the \ion{Ar}{i} and \ion{Si}{ii} column densities,
the $L/\ell$ ratio, the velocity dispersion $\sigma$, and the thermal widths
for each line, we can calculated their average absorption-line profiles
employing the generalized radiative transfer equation 
(Levshakov \& Kegel 1994, 1997).   
 
The same $\chi^2$-minimization 
procedure as before was applied to the same data set,
but now with two fitting parameters
$\{ N_{\ion{Si}{ii}}$, $\log ({\rm Ar}/{\rm Si})\}$, i.e. $\nu = 17$.
We fixed $\sigma = 13$ km~s$^{-1}$ and $T_{\rm kin} = 200$~K 
(i.e. $b \simeq 18$ km~s$^{-1}$) and calculated $\chi^2_{\rm min}$
for a given $L/\ell$. The results
are presented in Table~1. It should be emphasized that 
all \ion{Ar}{i} and \ion{Si}{ii} profiles
corresponding to the listed -- very different -- solutions are
{\it identical} to those shown in Fig.~1{\bf b}.

Our numerical results show that
accounting for correlation effects strongly affects the derived column densities
as well as the abundance ratio Ar/Si. These effects are closely related to
the changes in the curves of growths caused by the finite $L/\ell$ value.
Fig.~2 shows three curves of growth for each of the two lines. They correspond to
the two limiting cases of micro- and macroturbulence (Gail et al. 1974),
as well as to an intermediate -- mesoturbulent -- case ($L/\ell = 1.45$).
In general, one may say that the column density derived from
a measured equivalent width increases with increasing correlation length $\ell$.
Fig.~2 also allows a qualitative interpretation of the Ar/Si ratio in dependence of
$L/\ell$. In the microturbulent limit both lines are still close to the linear
part of the curve of growth, while in the mesoturbulent regime they lie on the
flat part. Since the slope here is smaller than on the linear part, the
Ar/Si ratio is larger. In the macroturbulent limit, both lines lie on the
square root part. In this case the slope is again steeper and, more importantly, the
two curves of growth are well separated from each other (due to the different
damping constants), implying a lower Ar/Si ratio than in the mesoturbulent case.
Thus, going from micro- to macroturbulence we expect the Ar/Si ratio at first
to rise and, after going through a maximum, to decline again. From Fig.~2 we
also see that in the macroturbulent limit the Ar/Si ratio is lower
than in the microturbulent one.

\section{Conclusions}

We have shown that metal abundances derived from
the absorption profiles of \ion{Ar}{i}\,$\lambda1048$ and
\ion{Si}{ii}\,$\lambda1020$, 
observed in the spectrum of the damped Ly$\alpha$ system
I~Zw~18 by Vidal-Madjar et al. (2000), 
become fully consistent with
those derived from 
emission-line 
spectra of the \ion{H}{ii} regions by Izotov \& Thuan (1999) and by
Izotov et al. (1999), 
if correlations in the large-scale velocity field are accounted for.
The adequacy of the obtained mesoturbulent solutions is supported by the fact
that the derived velocity
dispersion of the neutral gas, 
$\sigma \simeq 13$ km~s$^{-1}$, is in an excellent
agreement with the \ion{H}{i} 21~cm observations, $\sigma = 12 - 14$ km~s$^{-1}$
(van Zee et al. 1998).   
This implies that the gas in I~Zw~18 is efficiently mixed.  

To estimate more accurately metal abundances in the \ion{H}{i} gas within the
framework of our model, one needs to know the $L/\ell$ ratio.  
This parameter can be found from the analysis of saturated and 
optically thin lines of the same ion, since the latter are less affected
by the correlation effects. 
For our case, observations of the \ion{Ar}{i}\,$\lambda1066$
line would be of particular interest since its oscillator strength
$f_{1066} = 1/4\,f_{1048}$. The \ion{Ar}{i}\,$\lambda1066$ line was observed
in the $z = 3.4$ DLA system toward the quasar Q0000-2620 
and its analysis showed
a remarkably similar abundance to the other $\alpha$-chain elements 
O, S, and Si,
$Z/Z_\odot \simeq 1/80$ (Molaro et al. 2000).
The similarity of both DLA systems is also supported by the very low 
molecular hydrogen contents found in I~Zw~18, $f({\rm H}_2) 
\equiv 2N_{{\rm H}_2}/N_{\ion{H}{i}}
\ll 10^{-6}$ (Vidal-Madjar et al. 2000), and at $z=3.4$,
$f({\rm H}_2) \simeq 4\times10^{-8}$ 
(Levshakov et al. 2000).

Since Ar can hardly be depleted onto dust grains, 
but can be partially ionized by
nearby UV stellar radiation with energies
$h\nu > 15.76$ eV,
the relative abundance of \ion{Ar}{i} is a good indicator of the
intensity of the local photoionizing flux (Sofia \& Jenkins 1998).  
Our measurements rule out the presence of a significant amount of partially
ionized gas in the damped L$\alpha$ system I~Zw~18 and, hence, metal abundances
in the neutral gas do not require ionization corrections
as suggested by Izotov et al. (2000).

Thus, it is extremely important in future observations 
to investigate low-ion lines
to undestand better 
the kinematic characteristics of the neutral gas bulk motion  
which will enable us
to obtain more reliable estimations of metallicities.

\begin{acknowledgements}
The work of S.A.L. and I.I.A. is supported in part 
by the Deutsche Forschungsgemeinschaft and
by the RFBR grant No.~00-02-16007.
\end{acknowledgements}

\end{document}